\documentclass[preprint]{rmaa}

\title{Ionization-bounded and Density-bounded Planetary Nebulae}

 \author{Luis F. Rodr\'\i guez, Yolanda G\'omez, and Lizette Guzm\'an
  \affil{Centro de Radioastronom\'{\i}a y Astrof\'{\i}sica, UNAM, Morelia}
 }

\fulladdresses{
\item Yolanda G\'omez, Lizette Guzm\'an, and Luis F. Rodr\'\i guez: Centro de Radiostronom\'{\i}a
y Astrof\'\i sica, 
UNAM, A. P. 3-72, 
(Xangari), 58089 Morelia, Michoac\'an, M\'exico
(y.gomez; l.guzman; l.rodriguez@astrosmo.unam.mx).
}

\shortauthor{Rodr\'iguez et al.}
\shorttitle{IONIZATION-BOUNDED AND DENSITY-BOUNDED PN\lowercase{e}}

\SetVolume{50} \SetFirstPage{001} \SetYear{2009}
\ReceivedDate{\today} 
\AcceptedDate{Year Month Day} 

\resumen{Las nebulosas planetarias (y en general las regiones fotoionizadas)
pueden clasificarse como acotadas por ionizaci\'on (ionization-bounded) o
bien acotadas por densidad (density-bounded). 
Es importante poder determinar en que caso est\'a la nebulosa planetaria estudiada
para poder estimar de observaciones nebulares, por ejemplo, la tasa total de fotones ionizantes
producidos por la estrella central.
En este art\'\i culo presentamos un criterio observacional sencillo que utiliza
imagenes de radiocontinuo y que nos
permite establecer una condici\'on necesaria (pero no suficiente) para que la nebulosa planetaria
pueda ser considerada como acotada por ionizaci\'on. 
Aplicamos el criterio a dos nebulosas planetarias: NGC~7027
resulta muy posiblemente acotada por ionizaci\'on, mientras que Hb~4
est\'a acotada por densidad, al menos en ciertas direcciones.}

\abstract{
Planetary nebulae (and in general any photoionized region) can be classified as
ionization-bounded or density-bounded.
It is important to determine in which case is the planetary nebula studied to
be able to estimate from nebular observations, for example, the total rate of ionizing photons produced
by the central star. In this paper we present a simple observational 
criterion that uses radio continuum images and that allows to establish a
necessary (but not sufficient) condition for the planetary nebula
to be considered as ionization-bounded. We apply the criterion to two planetary nebulae:
NGC 7027 is most possibly ionization-bounded, while Hb~4 is density-bounded,
at least in some directions. 
}
      
\keywords{PLANETARY NEBULAE: INDIVIDUAL (NGC~7027, Hb~4) --- TECHNIQUES: IMAGE PROCESSING}

\begin{document}

\maketitle

\section{Introduction}

Planetary nebulae, and in general any type of nebula photoionized by a central
star or cluster of stars can be classified as either ionization-bounded or
density-bounded. In the first case, the radius of the ionized region is determined
by the absorption of the ionizing photons in the inner parts of the nebula.
Even if there is more external gas available, it will remain neutral.
In contrast, a density-bounded planetary nebula is one in which the ionized
region terminates simply because there is no more gas available. Even if there
are ionizing photons available, there will be no dense gas to ionize and the photons will
escape to the diffuse halos that are known to surround planetary
nebulae (Perinotto et al. 2004; Sandin et al. 2008).

When the nebula is visible in the optical,
the presence of lines such
as [O I] $\lambda$6300 (i. e. Bieging et al. 2007), that originate from the region of
transition between ionized and neutral hydrogen, suggests that the nebula is ionization-bounded. 
Also the association of the planetary nebula with  
neutral (Rodr\'\i guez \& Moran 1982) or molecular (Treffers et al. 1976) hydrogen or with
other molecular species (Thronson 1983) suggests ionization-boundness.
However, the situation is not as simple since a nebula can be ionization-bounded in
some directions (with respect to the ionizing star) 
and density-bounded in others and the evidence of a transition from ionized
to neutral or molecular gas could be coming from some directions only.

It is relevant to know if a nebulae is ionization-bounded in all directions since
in this case all the ionizing photon flux from the central star is trapped  
inside the nebula. Only in this case an accurate estimate of this ionizing flux can be derived
indirectly by observing the nebular emission, for example in recombination lines or free-free emission.
Otherwise, we will underestimate the ionizing photon flux.
In an important development, Zijlstra et al. (2008)
have reported Very Large Array
observations taken over 25 years of NGC~7027 at frequencies above 5 GHz, where the 
free-free emission is optically thin. They find that the
flux density is changing at a yearly rate of -0.145$\pm$0.005\%
and propose that this is caused by a decrease in the number of ionizing photons coming from the central star. 
However, their conclusion assumes that this planetary nebula is ionization-bounded in all directions.
This assumption seems very reasonable for NGC~7027 that is known to be associated with many 
molecular emissions and to possess large obscuration, suggesting that the ionized region is
completely engulfed in neutral gas. However, this assumption should be tested. 
Finally, Mellema (2004) and Sch{\"o}nberner et 
al. (2005) have noted that the apparent expansion velocity in the plane of the sky of
planetary nebulae that are ionization-bounded can be larger than the true expansion
velocity of the gas since we also have the contribution of the outward motion of the
ionization front as the nebula expands. This
discrepancy becomes quite large during the optically-thin
stage of nebular evolution (Sch{\"o}nberner et
al. 2005). The knowledge of whether or not a nebula
is ionization-bounded can be used to approximately correct for this effect.

\begin{figure}
\centering
\includegraphics[scale=0.70, angle=0]{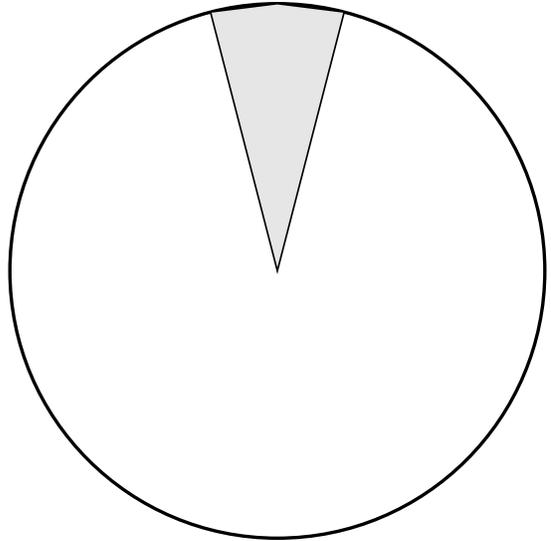}
 \caption{If we divide the emission from a nebula in the plane of
the sky in $n$ pie slices, each slice will have an opening angle
of $2 \pi / n$ radians (see figure) and its apex will coincide with the ionizing
star. From the point of view of an
observer in the central star, each of these slices will appear as
a spherical wedge with a solid angle of $4 \pi / n$ steradians.
}
  \label{fig1}
\end{figure}

In this paper we discuss a simple criterion that can be derived from
good quality radio continuum images taken at frequencies where the
ionized nebula is optically-thin. This analysis provides a necessary, although
not sufficient, condition to classify a nebula as ionization-bounded.
If the nebula fails the criterion, it can be considered to be density-bounded,
at least in some directions (as seen from the central star).

\section{The criterion}

We make the following assumptions: i) the central star produces a photoionizing flux that is
isotropic as seen from the position of the star, and ii) the nebula is optically-thin
in the free-free continuum. The first assumption implies that, if the nebula is
ionization-bounded, the rate of photoionizations (and recombinations) 
produced in the nebula is independent
of direction and is a constant for a given solid angle
(as seen from the star). Then, the number of free-free photons
(that in steady state is proportional to the number of recombinations, e. g. Schraml 
\& Mezger 1969), produced per differential
of solid angle as seen from the star is a constant. This conclusion is valid even if the radius of the
ionized gas is different in different directions. A larger density in a given direction
will cause a smaller ionized radius, but the number of free-free photons
produced per unit of volume will be larger, so the number of free-free photons
produced per differential of solid angle will remain constant. 
The second assumption implies that all free-free photons produced escape
from the nebula.

We now ask: what implications have these conclusions in the appearance of
the nebula in the sky? As seen by the observer, we can split the nebula
in $n$ identical ``pie'' slices centered in the central star (see Fig. 1),
each with an opening angle given by $2 \pi / n$ radians.  
As seen from the central star each of these slices is a spherical wedge with solid
angle given by $4 \pi / n$ steradians. Since this solid angle is constant for all
wedges, we conclude that for a nebula to be ionization-bounded, the flux density 
in each of the slices should be constant, within the noise.

This simple test can be implemented by following the next steps: 1) obtain a good quality
radio continuum image of a nebula at a frequency where it is known to be optically thin
in the free-free emission, 2) define a center for the nebula either using 
the stellar position or symmetry considerations applied to the nebula, 3)
use the task IRING of the NRAO software package AIPS
(Astronomical Image Processing System) to determine the flux density in each of $n$ identical slices
centered in the position determined in the previous step. The radius of these slices
has to be large enough to contain all the nebular emission. Finally, 4) plot these
flux densities as a function of the position angle of the slices to analyze 
their behavior.

\section{Observations}

The two sets of observations used in this study
were obtained in the standard continuum mode and
taken from the archive
of the VLA of the NRAO\footnote{The National Radio 
Astronomy Observatory is operated by Associated Universities 
Inc. under cooperative agreement with the National Science Foundation.}.
One set was for NGC~7027 ( = PN G084.9-03.4), and was taken at 43.34 GHz on 
2001 November 9 in the D configuration as part of the VLA Calibrator Flux Density project 
(R. A. Perley \& B. Butler 2008, in preparation).
The amplitude calibrator was
1331+305, with an adopted flux density of 1.46 Jy. 
NGC~7027 was used as its own phase calibrator,
with a bootstrapped flux density of 4.90$\pm$0.13 Jy.
This set was analyzed with the specific goal of testing
the hypothesis of ionization-boundness of
Zijlstra et al. (2008).
The second set was for Hb~4 ( = PN G003.1+02.9), and was taken at 8.46 GHz on 
2001 February 18 in the BnA as part of project AK528. The amplitude calibrator was
1331+305, with an adopted flux density of 5.21 Jy. 
The phase calibrator was 1745$-$290, 
with a bootstrapped flux density of 0.67$\pm$0.02 Jy.

In Table 1 we present, for each planetary nebula, the position adopted as the
center of the nebula (in both cases taken to be equal to the position of the minimum
emission in the ``bowl'' that characterizes the morphology of these objects),
and the frequency of the observations.

\begin{table}[htbp]
\small
  \setlength{\tabnotewidth}{0.45\textwidth} 
  \tablecols{4} 
  \caption{Observational Parameters}
  \begin{center}
    \begin{tabular}{lccc}\hline\hline
PN & $\alpha$(2000)$^a$ & $\delta$(2000)$^a$ & $\nu$(GHz)  \\
\hline
NGC~7027 & 21 07 01.74 & 42 14 09.4 & 43.34  \\
Hb~4 &  17 41 52.81 & -24 42 07.8 & 8.46  \\
\hline\hline
\tabnotetext{a}{Units of right ascension are hours, minutes, and seconds, and units of declination are
degrees, arcminutes, and arcseconds.}
    \label{tab:1}
    \end{tabular}
  \end{center}
\end{table}

The data were reduced using the standard VLA procedures in
the software package AIPS
of NRAO and then self-calibrated in phase and amplitude.
The images were made with the ROBUST weighting parameter
(Briggs 1995) of AIPS set to 0.

\section{Interpretation and Results}

\subsection{NGC~7027}

In Figure 2 we show the 43.34 GHz image of NGC~7027. This 
planetary nebula is known to be optically-thin above $\sim$5 GHz
(Zijlstra et al. 2008), so we expect this 43.34 GHz
image to be tracing optically-thin free-free emission.
In Figure 3 we plot the flux densities of each slice as a function of their
position angle. The slices were made with an angular
radius of 10$''$ and an opening of $10^\circ$, with the position angle
incrementing by $10^\circ$. This flux density per slice is approximately constant
(only small modulations at the $\sim$10\% level are present).
We conclude that this result is consistent with NGC~7027 being
ionization-bounded in all directions, in agreement with the assumption of Zijlstra et al. (2008).
This test is relevant because if NGC~7027 were density-bounded in some
directions, the observed decrease in flux density could be due to the
expansion of the nebula, allowing more ionizing photons to escape in the directions where
the nebula is density-bounded and not to a decrease in the number of ionizing 
photons from the central star. The assumption of NGC~7027 being ionization-bounded is
fundamental for the Zijlstra et al. (2008) analysis and our results are consistent with
this assumption.

\begin{figure}
\centering
\includegraphics[scale=0.45, angle=0]{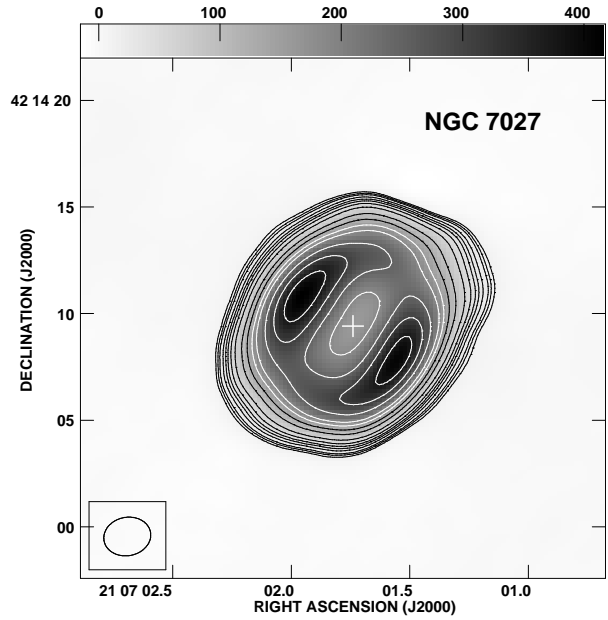}
 \caption{43.34 GHz contour and greyscale image of the planetary nebula NGC~7027.
The contours are -5, -4, 4, 5, 6, 8, 10, 12, 15, 20,
30, 40, 50, 60, 80, 100, and 120
times 3.1 mJy beam$^{-1}$, the rms noise of the image.
The greyscale is shown in the top wedge, in
mJy beam$^{-1}$.
The synthesized beam ($2\rlap.{''}21 \times 1\rlap.{''}80$
with a position angle of $-79^\circ$)
is shown in the bottom left corner
of the image. The cross marks the position adopted as the center of the nebular
emission (see Table 1).
}
  \label{fig2}
\end{figure}

\begin{figure}
\centering
\includegraphics[scale=0.40, angle=0]{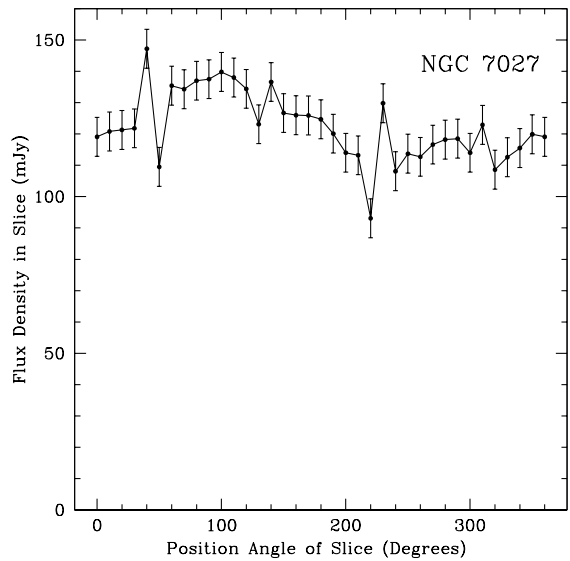}
 \caption{Flux density of the image slices of NGC~7027 as a function of their position angle.}
  \label{fig3}
\end{figure}

Unfortunately, the constancy in the flux density of the slices is
not a sufficient condition to assure ionization-boundness: an ionized ring
seen face-on would show this same constancy but it will be density-bounded along the
line of sight.

\subsection{Hb~4}

\begin{figure}
\centering
\includegraphics[scale=0.45, angle=0]{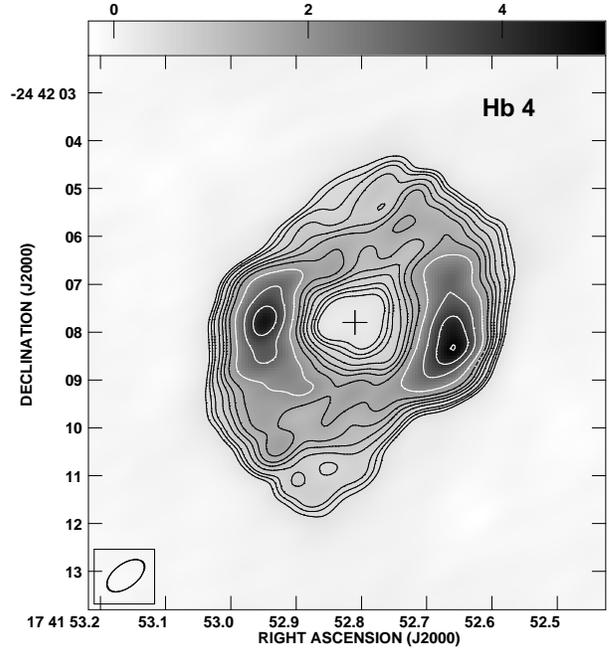}
 \caption{8.46 GHz contour and greyscale image of the planetary nebula Hb~4.
The contours are -5, -4, 4, 5, 6, 8, 10, 12, 15, 20,
30, 40, and 50
times 0.10 mJy beam$^{-1}$, the rms noise of the image.
The greyscale is shown in the top wedge, in
mJy beam$^{-1}$.
The synthesized beam ($0\rlap.{''}91 \times 0\rlap.{''}49$
with a position angle of $-53^\circ$)
is shown in the bottom left corner
of the image. The cross marks the position adopted as the center of the nebular
emission (see Table 1).
}
  \label{fig4}
\end{figure}

\begin{figure}
\centering
\includegraphics[scale=0.40, angle=0]{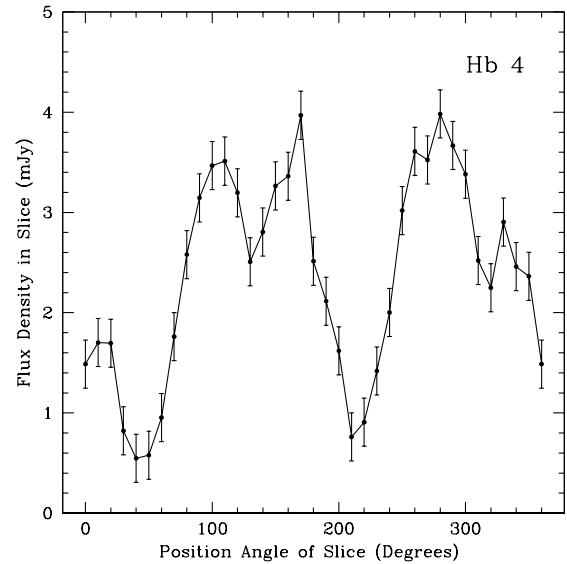}
 \caption{Flux density of the image slices of Hb~4 as a function of their position angle.} 
  \label{fig5}
\end{figure}

In Figure 4 we show the 8.46 GHz image of Hb~4. The 1.4 and 5.0 GHz flux densities of
this  planetary nebula are 158 mJy (Condon 
\& Kaplan 1998) and 170 mJy (Aaquist 
\& Kwok 1990), respectively.  This indicates that the nebula is 
optically-thin above $\sim$1.4 GHz,
so we expect the 8.46 GHz
image to be tracing optically-thin free-free emission.
In Figure 5 we plot the flux densities of each slice as a function of their
position angle. The slices were made, as for NGC~7027, with an angular radius
of 10$''$, every $10^\circ$. In contrast to NGC~7027, the flux density per slice
of Hb~4 shows a strong
modulation. This modulation implies that the nebula is density-bounded
at least in the NW and SE position angles, where ionizing photons will
be escaping to the interstellar medium. 
We conclude that this result indicates that Hb~4 is not ionization-bounded
in all directions. An estimate of the stellar ionizing flux from indicators
like H$\alpha$ or free-free emission will then result in an underestimate
of the true value. 

\section{Conclusions}

We  present a simple observational test that allows to establish
if an ionized nebula is ionization-bounded or not. The condition is
necessary but not sufficient. We applied this test to 
NGC~7027 and Hb~4, finding that the first object is probably
ionization-bounded while the second is density-bounded, at
least in some directions.


\acknowledgments
LFR and YG acknowledge the support
of DGAPA, UNAM, and of CONACyT (M\'exico).
This research has made use of the SIMBAD database, 
operated at CDS, Strasbourg, France.


\end{document}